\documentclass[twocolumn, prb,showpacs,aps]{revtex4}
\usepackage{graphicx}
\usepackage{amsmath}
\usepackage{epsfig}
\usepackage{color}
\usepackage{graphics}
\usepackage{latexsym}
\usepackage{amsfonts}
\usepackage{amssymb}
   
\usepackage{plain}

\begin{document}

\title{Spontaneous imbibition in disordered porous solids: a theoretical study of helium in silica aerogels}
\author{F. Leoni}
\affiliation{GIT-SPEC, CEA Saclay, 91191 Gif-sur-Yvette Cedex, France}
\author{E. Kierlik}
\affiliation{Laboratoire de Physique Th\'eorique de la Mati\`ere Condens\'ee, Universit\'e Pierre et Marie Curie, CNRS-UMR 7600\\ 4 place Jussieu, 75252 Paris Cedex 05, France}
\author{M. L. Rosinberg}
\affiliation{Laboratoire de Physique Th\'eorique de la Mati\`ere Condens\'ee, Universit\'e Pierre et Marie Curie, CNRS-UMR 7600\\ 4 place Jussieu, 75252 Paris Cedex 05, France}
\author{G. Tarjus}
\affiliation{Laboratoire de Physique Th\'eorique de la Mati\`ere Condens\'ee, Universit\'e Pierre et Marie Curie, CNRS-UMR 7600\\ 4 place Jussieu, 75252 Paris Cedex 05, France}



\begin{abstract}
We present a theoretical study of spontaneous imbibition of liquid $^4$He in silica aerogels focusing on the effect of porosity on the fluid dynamical behavior. We adopt a coarse-grained three-dimensional lattice-gas description like in previous studies of gas adsorption and capillary condensation, and use a dynamical mean-field theory, assuming that capillary disorder predominates over permeability disorder as in recent phase-field models of spontaneous imbibition.  Our results reveal a remarkable connection between imbibition and adsorption as also suggested by recent experiments. The imbibition front is always preceded by a precursor film and the classical Lucas-Washburn $\sqrt{t}$ scaling law is generally recovered, although some deviations may exist at large porosity. Moreover, the interface roughening is modified by wetting and confinement effects. Our results suggest that the interpretation of the recent experiments should be revised.

\end{abstract}

\maketitle

\def\be{\begin{equation}}
\def\ee{\end{equation}}
\def\bea{\begin{align}}
\def\eea{\end{align}}

\section{Introduction}

This paper is devoted to the theoretical description of spontaneous imbibition of helium in silica aerogels. It is motivated by recent experiments showing a striking influence of porosity and temperature on the fluid dynamical behavior\cite{N2006}. In particular, an intriguing two-step process has been observed in very light aerogels at low temperature, with two imbibition fronts rising successively inside the solid, both obeying the classical Lucas-Washburn (LW) linear relationship between penetration depth and square-root of time\cite{L1918}. Although these experiments have not yet been reproduced and cannot be easily interpreted at the microscopic scale, they  suggest the presence of precursor wetting films moving ahead of the primary front\cite{BQ2003}, as often observed in other systems, such as oil inside glass beads\cite{DC2010}, water in paper fibers\cite{AAEFD2008}, or during forced imbibition with immiscible fluids\cite{TMABCP1007}. More generally,  these experiments illustrate the subtle interplay between disorder, confinement, and wetting phenomena that takes place in aerogels and that is also at the origin of the  behavior observed in adsorption and desorption experiments\cite{WC1990,HDB2005,BLCGDP2008}. 
Silica aerogels are interesting materials for such studies because their porosity can be varied in a significant range and one can thus perform a systematic study of the influence of the microstructure on the fluid behavior. On the other hand, the heterogeneity of the gel at many scales and the absence of  a well-defined pore size make the dynamics more complex than in other disordered solids, such as Vycor glass\cite{HGSKV2007,G2010}. These features also greatly complicate the theoretical description.

Since spontaneous imbibition plays a crucial role in many problems of general interest (from oil recovery to agriculture) and is also used as a practical method to characterize porous structures (by modeling the solid as a bundle of uniform capillaries), it has received considerable attention over the years\cite{ADR2004}.   In spite of the diversity of actual porous media, the $\sqrt{t}$ scaling law, which results macroscopically from the balance between a constant driving capillary force and an increasing viscous drag, is ubiquitously observed. In fact, it appears that the LW relationship holds down to the nanoscopic scale, as shown by recent experimental\cite{HGSKV2007,G2010} and theoretical studies\cite{DMB2007,C2008,KPGPY2008,AK2009,WSP2010,SLCB2010}. This of course is an important indication for the development of nanofluidic devices. The dynamics of imbibition fronts is also a subject of much interest within the statistical physics community  as it is a good example of the propagation of an interface in a disordered medium (see Ref. \onlinecite{ADR2004} for a review and references therein).  Recent experimental\cite{GMH2002,SMPPR02005} and theoretical investigations\cite{ADR2004,DREAMA2000,PH2006} have especially focused on the roughening of the moving front that results from the random spatial distribution of capillary forces. It has been shown that the scaling properties of the interface are strongly affected by the non-locality induced by fluid conservation. Because a microscopic treatment of liquid flow in a random medium is a formidable problem,  these theoretical approaches generally use a coarse-grained description in which the spatial configurations of the fluid are represented by a phase-field model\cite{DREAMA2000,PH2006}    and the disordered structure of the solid is represented by quenched, spatially uncorrelated random fields.  While this description is appropriate to investigate the large-scale properties of the interface, it does not account for the correlations in the solid microstructure (for instance along the silica strands in aerogels) nor for the wetting layers that may be present on the pore walls, depending on the solid wettability. In order to interpret the imbibition experiments in aerogels\cite{N2006}, one therefore needs a somewhat more involved  (though still coarse-grained) description in which the solid microstructure is taken into account, and the temperature, the porosity, and the solid-fluid interaction can be varied independently.

In recent years, such a description, based on a lattice-gas model and a mean-field statistical treatment, has been used  to investigate the hysteretic capillary condensation of gas in disordered porous solids\cite{KMRST2001,WSM2001,RKT2003}. In particular, the experimental results in aerogels\cite{WC1990,HDB2005,BLCGDP2008} have been faithfully reproduced, and the remarkable dependence of the adsorption isotherms on temperature and porosity has been elucidated\cite{DKRT2003,DKRT2004,DKRT2005}. In the present paper,  we apply the same model to spontaneous imbibition, using a dynamical version of lattice mean-field theory\cite{GPDM2003} that has already proven useful to model condensation and evaporation processes in nanopores\cite{M2008,EM2009}. Like in the phase-field models of imbibition\cite{DREAMA2000,PH2006}, we consider the regime where capillary disorder predominates over permeability disorder, which corresponds to the case of a slowly advancing interface. Only diffusion-like mechanisms are thus included and the effect of the hydrodynamic modes are incorporated through effective parameters. However, the key ingredient at the origin of the constant slowing down of the imbibition front predicted by the LW law, namely liquid conservation, is explicitly treated. In a previous work\cite{KLRT2010}, we have applied this theory to spontaneous imbibition in a single slit-pore, emphasizing the role of precursor wetting films on the propagation of the main meniscus.  This was a first necessary step before addressing the much more complex case of disordered solids.

The paper is organized as follows. In the next section, we review the lattice gas model and the dynamical mean-field theory. The numerical results are presented and discussed in Sec. III. Finally, in Sec. IV, we summarize our findings and conclude. Some details about the definition of a local, disorder-dependent permeability are given in the Appendix. 

\section{Model and theory}

As described in detail in  Refs. \onlinecite{DKRT2003,DKRT2004,DKRT2005}, we model the gel-fluid system by a coarse-grained nearest-neighbor lattice gas with a configurational energy function given by 
\begin{equation}
\label{eq:hamilton}
\mathcal{H}=-w_{ff}\sum_{\langle ij\rangle}\tau_i\tau_j\eta_i\eta_j-w_{sf}\sum_{\langle ij\rangle}[\tau_i\eta_i(1-\eta_j)+\tau_j\eta_j(1-\eta_i)]
\end{equation}
where $\tau_i=0,1$ is the fluid occupancy variable at site i ($i=1,...,N$) and $\eta_i=1,0$ is a quenched random variable that describes the solid microstructure ($\eta_i=0$ if site $i$ is occupied by a gel particle); $w_{ff}$ and $w_{sf}$ denote the fluid-fluid and solid-fluid
attractive interactions, respectively, and the sums run over all distinct pairs of nearest-neighbor (n.n.) sites. The ratio $\alpha=w_{sf}/w_{ff}$ controls the wettability of the solid surface and $\phi=(1/N)\sum_i \eta_i$ defines the solid porosity. 

Gel configurations ({\it i.e.}, the set of $\eta_i$'s) are generated by a standard on-lattice diffusion-limited cluster-cluster aggregation (DLCA) algorithm\cite{M1983} which has been shown to faithfully represent the structure of the base-catalyzed silica aerogels which are considered here.  A body-centered-cubic (bcc) lattice is used in order to avoid some lattice artifacts at low temperature\cite{DKRT2004}.  From the computation of the gel correlation length $\xi_G$, the lattice constant $a$  (hereafter taken as the unit length) was estimated in Ref. \onlinecite{DKRT2003} to be of the order of a few nanometers, which is consistent with the coarse-grained picture of a gel site representing a SiO$_2$ particle with a diameter of about $30$\AA.
The lattice has lateral size $L_x=L_y=L$ in the $x$ and $y$ directions and height $L_z$ in the $z$ direction ({\it i.e.}, $N=2L_zL^2$)\cite{note1}. Periodic boundary conditions are taken in the $x$ and $y$ directions and the liquid reservoir is located at the bottom of the simulation box, as illustrated in Fig. 1.

\begin{figure}[hbt]
\begin{center}
\label{Fig1}
\includegraphics[width=7cm]{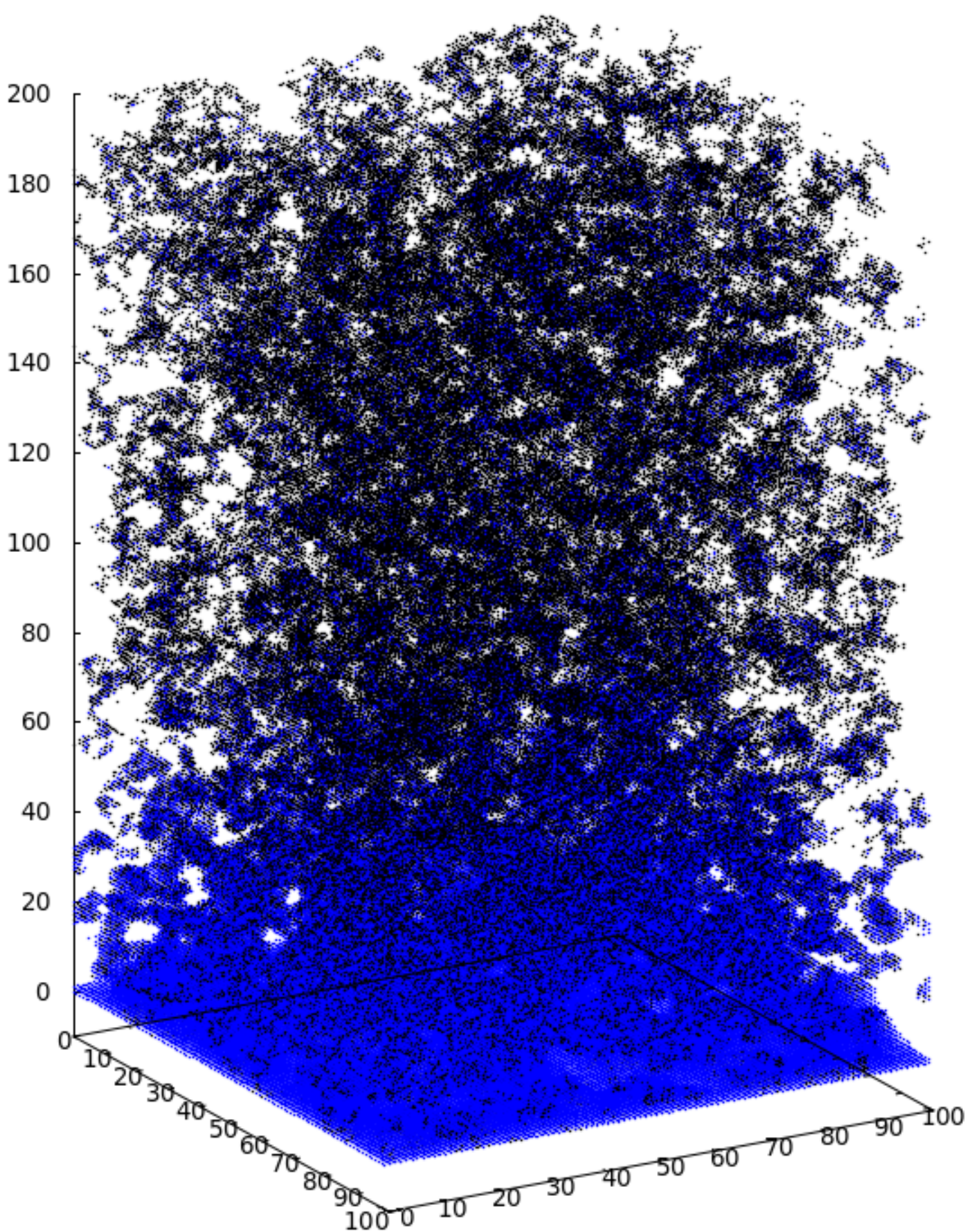}
\caption{(Color on line) Snapshot of liquid configurations during the imbibition process in a $95$\% porosity DLCA lattice aerogel ($L_x=L_y=100, L_z=200$) at time $t=30$.  Fluid and gel particles are coded in blue and black, respectively, and only the fluid sites with a density $\rho_i$ larger than $0.5$ are shown. The liquid reservoir in located at the bottom of the simulation box.}
\end{center}
\end{figure}

The theoretical approach that we use to model the liquid dynamics may be viewed as a mean-field version of the Kawasaki spin exchange dynamics and was originally applied to study phase separation and surface enrichment in solid binary alloys\cite{B1974}  (see Ref. \onlinecite{GPDM2003} for  a comprehensive review, Ref. \onlinecite{MAD2004} for an application to diffusion in membranes and bulk fluids and Refs. \onlinecite{M2008} and \onlinecite{EM2009} for a recent application to fluids confined in nanopores). It is referred to as ``mean-field kinetic theory" or ``dynamic mean-field theory "(DMFT). The starting point is the exact master equation describing the temporal evolution of $\rho_i(t)=\langle \tau_i \eta_i\rangle_t$, the ensemble average fluid density at site $i$. Assuming that transport is due to a hopping process between nearest-neighbor sites and replacing the occupancy variables by their ensemble average ({\it i.e.} neglecting dynamical correlations), one can then derive a continuity equation describing the local conservation of mass 
\begin{equation}
\label{Eqevol}
\frac{\partial\rho_i(t)}{\partial t}+\sum_{j/i}J_{ij}(t)=0
\end{equation}
where the sum runs over all n.n. of site $i$ and the flux $J_{ij}(t)$ is given by 
\begin{equation}
\label{Eqflux}
J_{ij}(t)=w_{ij}\rho_i(\eta_j-\rho_j)-w_{ji}\rho_j(\eta_i-\rho_i) 
\end{equation}
(note the presence of the random variables $\eta_i$ and $\eta_j$ that describe the configuration of the gel particles on the lattice). The $w_{ij}$'s are the Metropolis transition probabilities associated to the Kawasaki dynamics in the mean field approximation,
\begin{equation}
\label{wij}
w_{ij}(\{\rho\})=w_0\exp(-\beta E_{ij})
\end{equation}
where
\begin{equation}
E_{ij}=\left\{
\begin{array}{ll}
0,       & E_j<E_i\\
E_j-E_i, & E_j>E_i
\end{array}\right.
\end{equation}
and
\begin{equation}
\label{Eqener}
E_i=-w_{ff} \sum_{ j/i}\rho_j-w_{sf}\sum_{j/i}(1-\eta_j)
\end{equation}
as obtained from Eq. (1); $w_0$ is an elementary jump rate that sets the time scale and is considered here as an effective parameter incorporating some information about the hydrodynamic modes that are not  explicitly treated. It is indeed clear that this approximate treatment of the actual dynamics is essentially diffusive, in the same way as the recent mesoscopic phase-field models of spontaneous and forced-flow imbibition\cite{ADR2004,DREAMA2000,PH2006}. The analogy with a continuum phase-field description becomes more apparent if the continuity equation is recast into the discrete version of a generalized Cahn-Hilliard equation\cite{CH1958}, or a dynamic density functional theory\cite{MT1999,AE2004}, introducing an effective mobility coefficient  related to the local site densities via the Metropolis transition probabilities\cite{M2008} (see in particular the appendix of Ref. \onlinecite{KLRT2010} where the continuum limit of the above equations is derived in the much simpler case of a single slit pore). 

The connection of Eq. (\ref{Eqevol}) to the phenomenological Darcy's equation\cite{ADR2004} that relates the flux of liquid in a porous medium to the macroscopic pressure gradient, 
\begin{equation}
{\bf J}=-\frac{\kappa}{\eta}  \boldsymbol{\nabla} p \ ,
\end{equation}
is established by taking $w_0$ proportional to the ratio $\kappa/\eta$ of the solid permeability to the fluid viscosity (as in the phase-field models\cite{ADR2004}). The permeability $\kappa$ represents the (volume-averaged) frictional effects  exerted by the solid microstructure on the imbibing  fluid and is essentially dependent on the size of the pores through which the liquid flows (for a single capillary of radius $R$,  $\kappa \propto R^2$ if one assumes a Hagen-Poiseuille laminar flow). For porous solids with a reasonable homogeneous cylindrical pore structure such as Vycor, $\kappa$ can thus be related to phenomenological parameters characterizing the solid morphology such as the porosity $\phi$, the average pore radius $r_0$, or the tortuosity $\tau$\cite{HGSKV2007,G2010}. This relationship is much more elusive in aerogels since the structure is inhomogeneous at many scales, as already emphasized.  However, in order to compare results obtained in aerogels with different porosities, it is crucial to take into account some dependence of $w_0$ on the solid properties. In a first approximation, one may simply assume that $\kappa$, and consequently $w_0$, only depends on $\phi$, neglecting local variations   due to the disorder of the gel structure. Several theoretical expressions relating $\kappa$ to $\phi$ are available in the literature\cite{KS2002}. In particular, Brinkman's effective-medium theory\cite{B1947} appears to be a good representation of the flow  in a dilute, disordered porous matrix down to the nanoscale, as shown by very recent lattice-Boltzmann and molecular simulations\cite{RHSS2010}.  For a dilute collection of non-overlapping spheres of radius $R$ and volume density $n$ (so that the volume fraction is $1-\phi=4\pi R^3n/3$), Brinkman's expression for $\kappa(\phi)$ is 
\begin{equation}
\label{EqBrink}
\kappa(\phi)=\kappa_0\left[1+\frac{3}{4}(1-\phi)\left(1-\sqrt{\frac{8}{1-\phi}-3}\right)
\right] 
\end{equation}
where  $\kappa_0=1/(6\pi Rn)=2R^2/[9(1-\phi)]$ is the permeability in the high-dilution limit\cite{LL1959}. Brinkman's theory somewhat underestimates the permeability of a dilute gel network\cite{RHSS2010} but the correction is almost independent of $\phi$ in the regime considered in the present study ($\phi \ge 87\%$).
The transposition to the case of a discrete b.c.c. lattice suggests to take $R/a=\sqrt{3}/4$ so that two spheres representing silica particles and placed at n.n. sites are tangent (this choice, however, is not essential in our treatment since the value of $R$ is the same for all aerogels). This yields the values indicated in Table 1. Overall, assuming that $w_0$ has the same dependence on $\phi$ as the permeability  simply amounts to replacing the imbibition time $t$ in Eq. (\ref{Eqevol}) by a rescaled, porosity-dependent, time $t(\phi)=t/\kappa(\phi)$. \begin{table}[htbp]
\begin{center}
\begin{tabular}{|c|c|}
\hline
\ \ \ \ $\phi$ \ \ \ \ & \ \ \ \ $\kappa(\phi)$ \ \ \ \ \\
\hline
\hline
$0.87$ & 0.113 \\
\hline
$0.90$ &  0.174\\
\hline
$0.92$ & 0.244 \\
\hline
$0.95$ & 0.473 \\
\hline
\end{tabular}
\end{center}
\caption{Permeability $\kappa(\phi)$ computed from Eq. (\ref{EqBrink}) with $R/a=\sqrt{3}/4$.}
\label{tab:table}
\end{table}\\

Of course, this is only an average description and one would like to also take into account  the influence of local inhomogeneities in the microstructure.  This can be done for instance by defining a local porosity obtained by averaging the number of gel particles in a spherical region around each fluid site. Such an approximate treatment, which introduces some disorder into the permeability (hence into $w_0$), is described in the Appendix. It turns out, however, that this correction is almost negligible. Indeed, as emphasized in Refs. \onlinecite{ADR2004,DREAMA2000} in the framework of phase-field models, the effect of disorder on the mobility is essentially proportional to the interfacial velocity, which is never large in the case of spontaneous imbibition and moreover continuously slows down (as $t^{-1/2}$ according to the macroscopic LW law). Therefore, disorder-induced fluctuations of the capillary forces should prevail over fluctuations of the permeability, and the former are properly taken into account in our treatment via the dependence of local energy $E_i$  on the quenched variables $\eta_i$ representing the solid microstructure. 

Finally, it is important to observe that the local densities $\{\rho_i\}$ minimizing the static (mean-field) Helmholtz free energy  ${\cal F}[\{\rho_i\}]$ in the canonical ensemble 
\begin{align}
\label{Freeener}
&\beta {\cal F}[\{\rho_i\}]=\sum_{i} [\rho_i \ln (\rho_i) + (\eta_i-\rho_i)\ln (\eta_i-\rho_i)] \nonumber\\
&-\beta w_{ff} \sum_{<ij>} \rho_i\rho_j -\beta w_{sf}  \sum_{<ij>} [ \rho_i(1-\eta_j) + \rho_j(1-\eta_i)] 
\end{align}
(with $\beta=1/k_BT$) or the static (mean-field) grand-potential $\Omega = {\cal F} -\mu \sum_i \rho_i$ in the grand canonical ensemble, are steady-state solutions of the DMFT equations\cite{M2008}. Indeed, one can readily check that the $\rho_i$'s  satisfying the set of coupled equations $\partial {\cal F}/\partial \rho_i=\mu$, {\it i.e.}
\begin{equation}
\label{Eqmin}
\frac{\rho_i}{\eta_i-\rho_i}=\exp [\beta (\mu-E_i)] \ ,
\end{equation}
are solutions of Eq. (\ref{Eqevol}) when $\partial \rho_i(t)/\partial t=0$. This feature will allow us to unveil a remarkable connection between the microscopic states visited dynamically during the imbibition process and those visited during quasi-static adsorption experiments performed in the same systems. In a disordered material, Eq. (\ref{Eqmin}) has many solutions in general, typically of the order of $\exp(N)$, and these local minima\cite{note1a}  correspond to metastable states whose presence underlies the out-of-equilibrium behavior observed in adsorption/desorption experiments\cite{KMRST2001,WSM2001,RKT2003,DKRT2003,DKRT2004,DKRT2005}. This feature also plays a crucial role in imbibition, as the interface may be pinned in one of these states, resulting in an intermittent, avalanche-like motion\cite{ADR2004}.

In practice, we have integrated Eq. (\ref{Eqevol}) via Euler's method [{\it i.e.}, $\rho_i(t+\Delta t)=\rho_i(t)-\Delta t\sum_{j/i}J_{ij}(t)$] by using the dimensionless time step  $w_0(\phi) \Delta t=0.1$. This value was small enough to ensure the stability of the solution in all cases. However, it corresponds to a very small variation of the local site densities. This was already found in our previous study of imbibition in a single slit-pore\cite{KLRT2010},  but the present fully three-dimensional computation is considerably more time-consuming. As a result, the penetration depth of the main interface remains small compared to the overall height of the simulation box, even after several months of CPU time. This is a serious technical problem which, unfortunately, seems to be inherent to this type of approach. Despite a great deal of effort, we have not found an alternative method. For plotting  the numerical results, we use a larger time scale, $t_0= 10^4w_0 \Delta t$, which is hereafter taken as the effective time unit.

\section{Results and discussion}

We have studied aerogels with porosity $\phi=87\%$, $90\%$, $92\%$, and $95\%$ but  for the sake of brevity we mainly focus on the $87\%$ and $95\%$ solids in the following. The results correspond to the case of a primary imbibition, when the liquid in the reservoir is initially in contact with the porous solid filled by (bulk) vapor at $\mu = \mu_{sat}$\cite{note2} (this is not an equilibrium condition due to the presence of the porous solid). This is realized by fixing the density of the fluid at the saturated liquid density $\rho_l(T)$ in the first layer at the bottom of the box ($z=0$)  and at the saturated vapor density $\rho_g(T)$ in the rest of the gel. Then, at $t=0$,  we let the system evolve according to Eq. (\ref{Eqevol}) and the liquid enters the gel. 

 To make contact with previous theoretical studies of $^4$He adsorption\cite{DKRT2003}, the interaction ratio $\alpha=w_{sf}/w_{ff}$ is set equal to 2 in order to reproduce approximately the height of the sorption hysteresis loop measured  in a $87\%$ aerogel at $T=2.42$K.   However, to prevent the liquid from escaping, a drying condition is imposed in the top layer  by setting $\alpha=0$ for $z=L_z$.  All calculations are performed at the same reduced temperature $T^*=k_BT/w_{ff}=0.9$ ({\it i.e.} $T/T_c=0.45$) so that $\rho_g\approx 0.013$ and $\rho_l\approx 0.987$.  

\begin{figure}[hbt]
\begin{center}
\includegraphics[width=8cm]{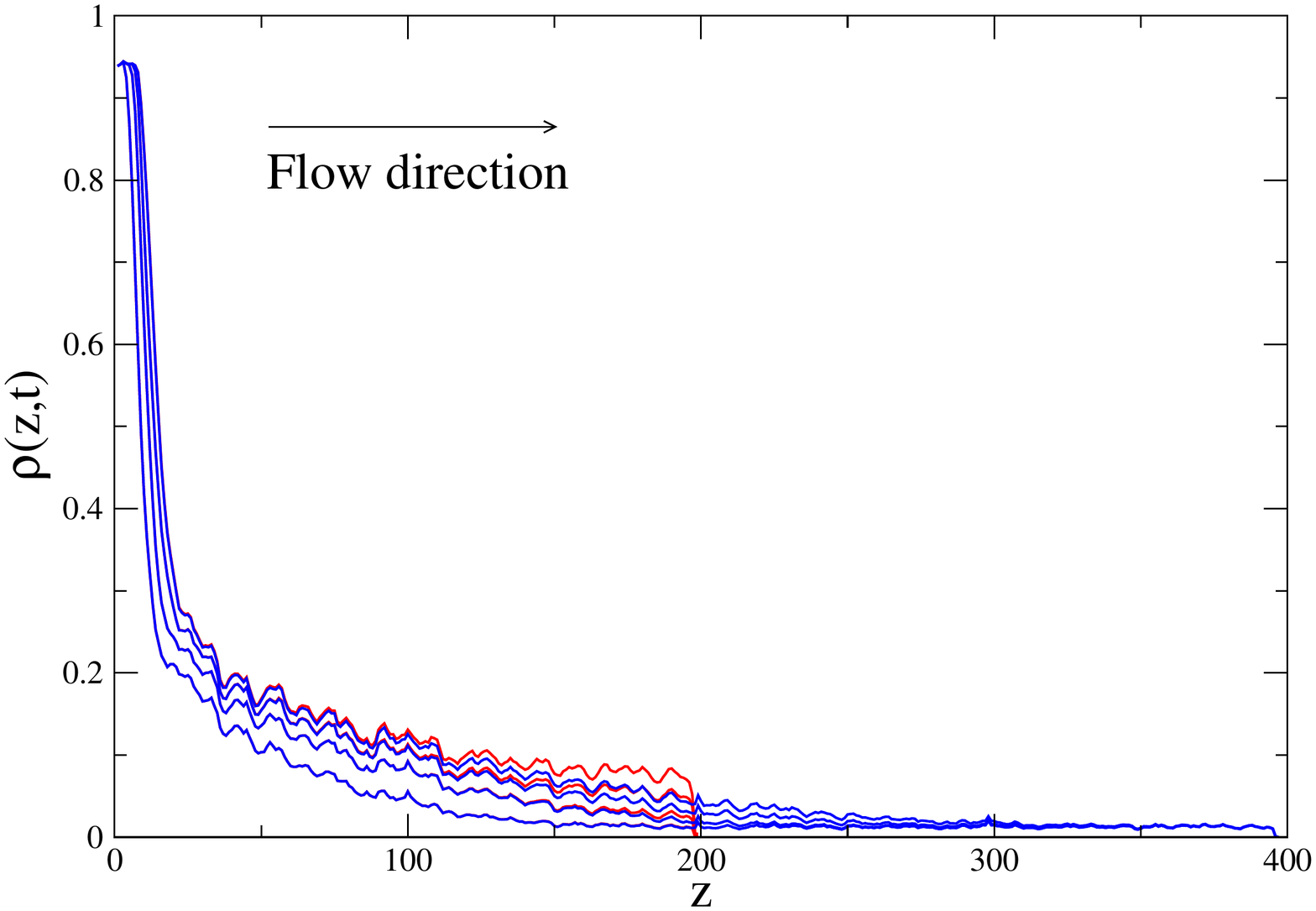}
\caption{(Color on line) Profiles of the average fluid density $\rho(z,t)$ at different imbibition times ($t=30,60,90, 120$ from bottom to top) in two $95$\% aerogels with sizes  $L_z=2L=200$ (red lines) and $L_z=4L=400$ (blue lines). The latter sample has been duplicated from the former\cite{note1}. The length and time units are $a$ and $t_0$, respectively.}
\label{Fig2}
\end{center}
\end{figure}
As emphasized in Ref. \onlinecite{KLRT2010}, the possible presence of precursor films that advance faster than the main interface and that quickly reach the other end of the porous solid makes the system size $L_z$ in the direction of the flow a critical parameter in this type of study\cite{note3}.
This is illustrated in Fig. \ref{Fig2}  where we plot the time evolution of the fluid density profile $\rho(z,t)$ obtained by averaging the density in the $x-y$ plane and we compare the results for two samples of a $95\%$ porosity aerogel with sizes  $L_z=2L=200$ and $L_z=4L=400$. 
The characteristic features of the profile will be discussed in more detail below but one can see at once that because of the boundary condition imposed at $z=L_z$  the fluid accumulates at the top end of the solid, which induces an artificial thickening of the density profile  (this is not due to a ``resistance" of the gas occupying the gel during the liquid invasion\cite{DBCPTS2009} since the gas can always condense). However, since it takes some time for the precursor film to reach the other end of the solid (see below), these finite-size effects can be neglected so long as the imbibition time $t$ is shorter than some time $t_{max}(z)$  (the larger $z$, the smaller $t_{max}$). For instance, in Fig. \ref{Fig2}, the profiles obtained with $L_z=200$ are insensitive to the boundary effects if the study is restricted to $z\lesssim100$ and $t\lesssim120$. If one is only interested in the small-$z$ region, typically $z\lesssim20$, larger times may be considered, up to  $t\approx 200$. The results that we discuss hereafter were obtained  with a system of height $L_z=2L=200$ (whence $N=4\times10^6$).
\begin{figure}[hbt]
\begin{center}
\label{Fig3}
\includegraphics[height=9.5cm,width=9cm]{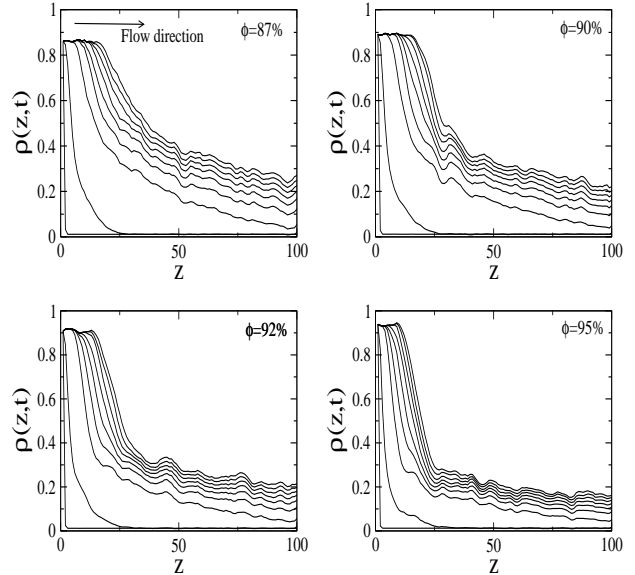}
\caption{Density profiles $\rho(z,t)$ in aerogels with porosity $87\%$, $90\%, 92\%$ and $95\%$  at times $t=1, 2, 30, 60, 90, 120...210.$ The data are averaged over $3$ gel realizations for $\phi=87\%$ and $5$ realizations for the other porosities.}
\end{center}
\end{figure}

To begin with, we show in Fig. 3  the time evolution of the profiles $\rho(z,t)$ in the different aerogels, after averaging the numerical data over a small number of gel realizations to somewhat smooth out the fluctuations over disorder (it was not possible to consider more realizations because of the huge computational time required). Note that the time unit is $t_0$ (just proportional to the number of time steps in the integration of Eq. (\ref{Eqevol}) by Euler's method) and is therefore not renormalized  by the porosity-dependent factor $\kappa(\phi)$ given by Eq. (\ref{EqBrink}). We are indeed only interested here in observing qualitatively the influence of porosity on the density profiles and not in comparing the imbibition rates (as will be shown below, the imbibition front actually moves faster as the porosity increases when time is properly rescaled). The most salient feature in all cases is that the density is much larger than $\rho_g$ far ahead of the main interface, indicating the presence of a precursor ``wetting" film that thickens with time. Whereas the imbibition front cannot be clearly discriminated from the precursor film for $\phi=87\%$, there are clearly two distinct regions in the profile for $\phi=95\%$. In this latter case, the arrival of the imbibition front at a given distance $z$ from the liquid reservoir is signaled by a steep increase of the density from about $\rho\approx 0.30$ to $\rho\approx \rho_l (1-\phi)=0.94$. Note that the precursor has already reached the middle of the sample at $t\approx 30$  while the front is still located at $z \approx 10$ (see Fig. 3).

\begin{figure}[hbt]
\begin{center}
\includegraphics[width=9cm]{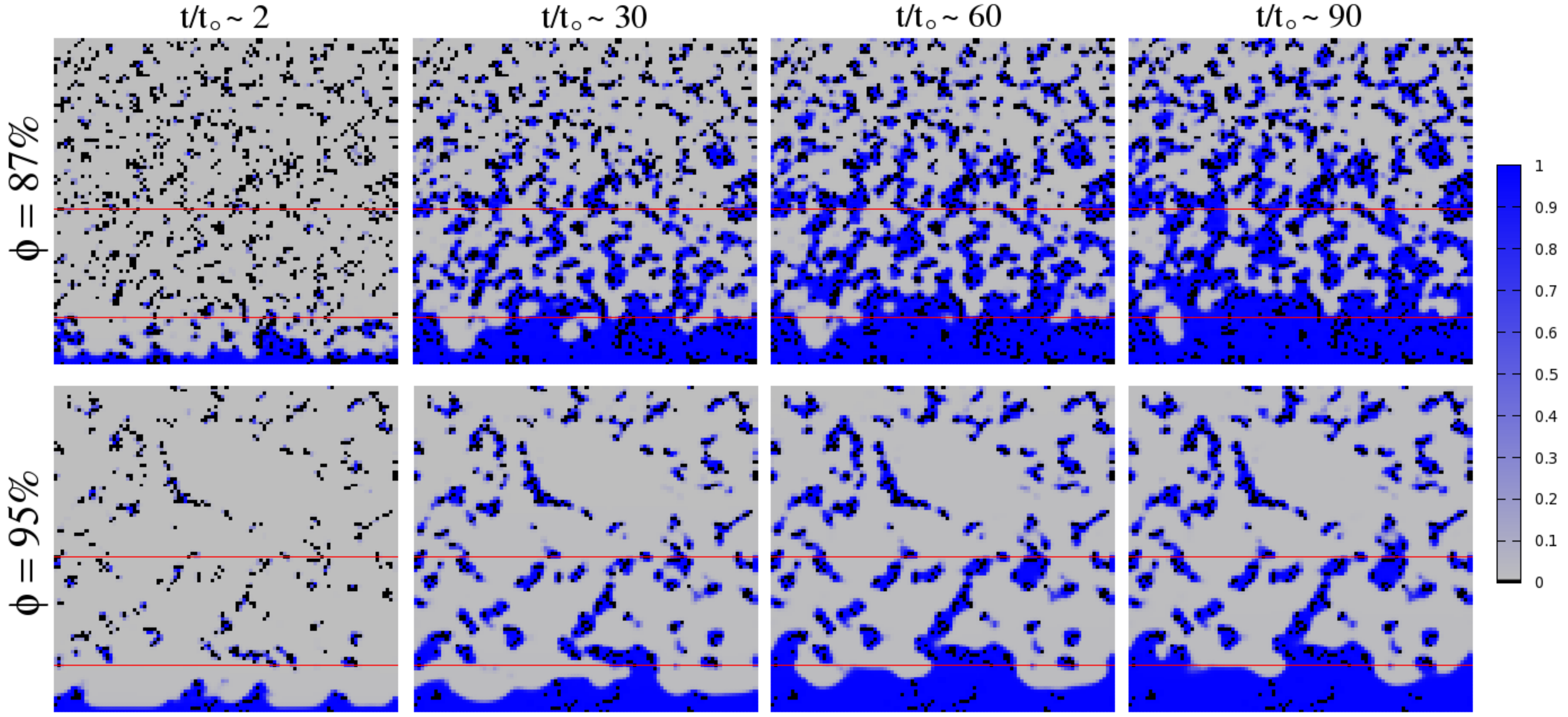}
 \caption{(Color on line) Snapshots of the liquid configurations in $87\%$ (top) and $95\%$ (bottom) aerogel samples taken in the vertical plane $x=L/2=50$ (only the region $z< L$ is shown) at different times. The aerogel particles are shown in black and the fluid density is represented with the color scale indicated in the figure  (from $\rho_i\approx0$ in grey to $\rho_i=1$ in dark blue). The red lines indicate the position of the horizontal planes corresponding to the cross sections shown in Figs. \ref{Fig5} and \ref{Fig6}.}
 \label{Fig4}
\end{center}
\end{figure}

In order to better visualize the underlying microscopic mechanisms, we show in Fig. \ref{Fig4} some typical snapshots of the liquid configurations in two samples of $87\%$ and $95\%$ porosity. These snapshots are taken at successive times in a vertical plane located in the middle of the solid ($x=L/2=50$).

\begin{figure}[hbt]
\begin{center}
\includegraphics[width=9cm]{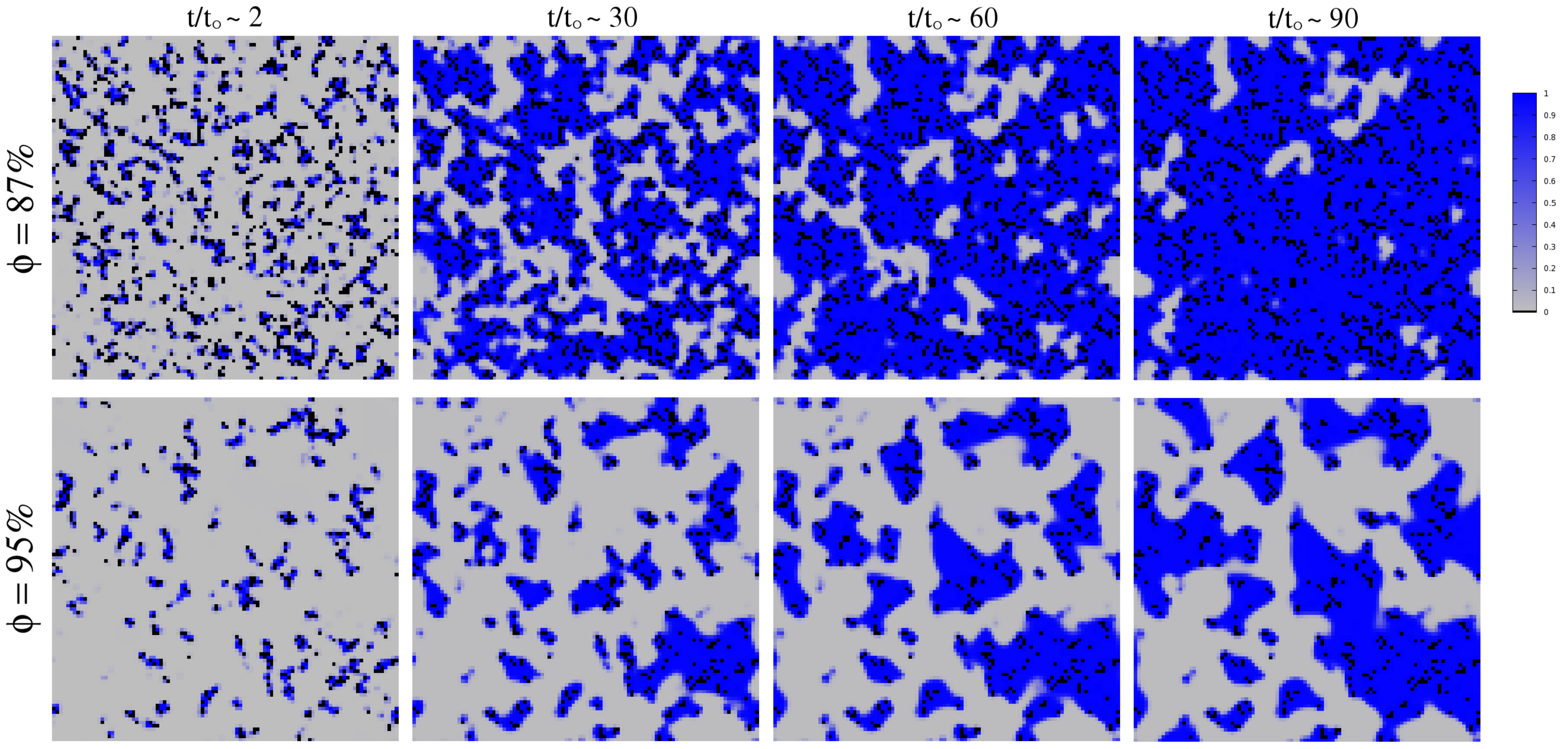}
 \caption{(Color on line) Horizontal cross sections in the plane $z=15$ at different times $t$. The samples are the same as in Fig. \ref{Fig4}. }
\label{Fig5}
\end{center}
\end{figure}

\begin{figure}[hbt]
\begin{center}
\includegraphics[width=9cm]{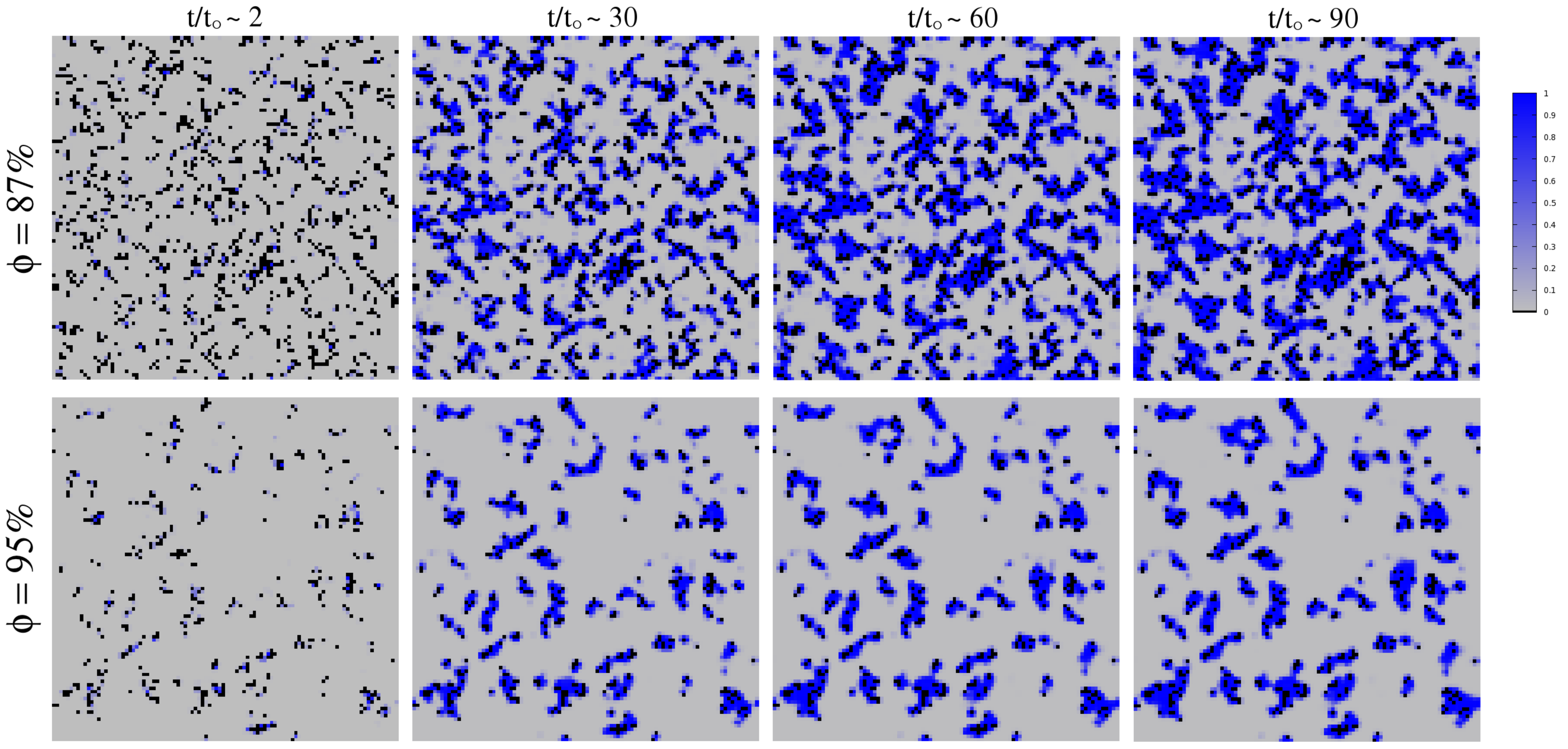}
 \caption{(Color on line) Same as Fig. \ref{Fig5} for $z=50$. }
\label{Fig6}
\end{center}
\end{figure}

 One can see that the fluid density distribution is essentially bimodal at this temperature ({\it i.e.}, $\rho_i\approx \rho_g$ or $\rho_i\approx \rho_l$) with a negligible fraction of sites with an intermediate density. This feature was also observed in the study of helium adsorption in aerogels\cite{DKRT2003}. In fact, the configurations of the imbibed liquid closely resemble those observed during adsorption as one slowly varies the chemical potential (or the pressure) in the external reservoir\cite{DKRT2003,DKRT2004,DKRT2005}. 
This is even more visible in Figs. \ref{Fig5} and \ref{Fig6} that display horizontal cross sections of the system at two distances $z$ from the liquid reservoir (which is somewhat equivalent to examine the same cross section at successive times: in this sense, Fig. \ref{Fig6} is a zoom on the first stages of Fig. \ref{Fig5}).  Like in adsorption, the first stage of the imbibition process is a coating of the silica strands by a liquid film. (In fact, as soon as the system is let to evolve, there is a quasi-instantaneous redistribution of the gas molecules due to the attraction exerted by the solid, as can be seen for instance in Fig. \ref{Fig6} for $t = 2$; this transient process determines the true initial configuration of the fluid inside the porous solid and has nothing to do with the arrival of the liquid from the reservoir\cite{KLRT2010}.) The wetting film then progressively thickens while the voids or crevices between the aerogel strands fill with liquid. During this second stage, the porosity of the solid plays a crucial role since the size and shape of the voids are quite different in the two aerogels\cite{DKRT2003,DKRT2004,DKRT2005}. This has also an important consequence on the morphology and the advance of the imbibition front, as can be seen in Fig. \ref{Fig4}. 

\begin{figure}[hbt]
\begin{center}
\includegraphics[width=8cm]{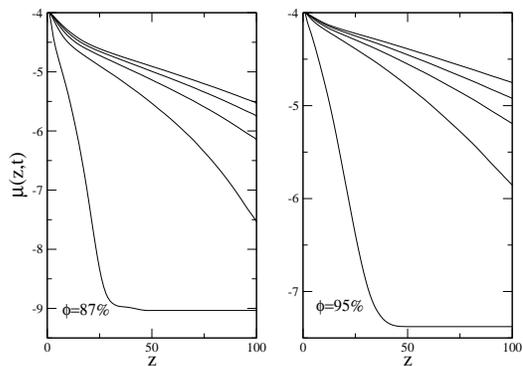}
 \caption{Chemical potential profile $\mu(z,t)$ in the $87\%$ and $95\%$ aerogels at $t=1,30,60,90,120$ (from left to right). }
\label{Fig7}
\end{center}
\end{figure}

When looking at the density profiles in the $87\%$ and $95\%$ aerogels in Fig. 3, one cannot help but notice a similarity with the corresponding adsorption isotherms computed in Refs. \onlinecite{DKRT2003,DKRT2004}. Those were obtained from Eqs. (\ref{Eqmin}) by starting  from a low chemical potential and increasing $\mu$ in small steps.  As also observed experimentally, these isotherms are smooth in the $87$\% aerogel but become much steeper in the $95$\% one (the theory actually predicts a genuine jump in the average density at low enough temperature). To check if there exists a more quantitative relationship between the two processes, it is instructive to compute the (dynamical) chemical potential profile $\mu(z,t)$ defined as the average of the local chemical potential $\mu_i(t)=k_BT \ln[\rho_i(t)/(1-\rho_i(t))] +E_i(t)$ over the transverse directions. Some profiles $\mu(z)$ at different imbibition times are shown in Fig. \ref{Fig7}. 

\begin{figure}[hbt]
\begin{center}
\includegraphics[width=8cm]{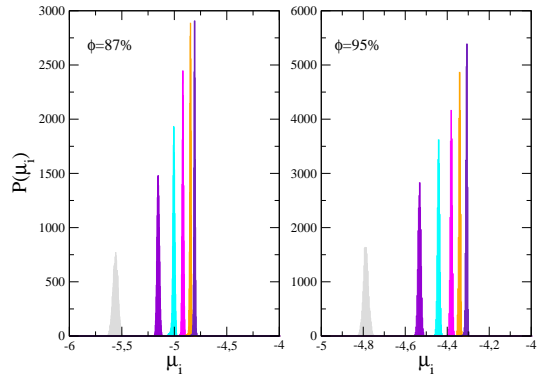}
 \caption{(Color on line) Histograms of the local chemical potentials $\mu_i$ in the $87\%$ and $95\%$ aerogels at $z=50$ and $t=30,60,90,120,150,180$ (from left to right). The small dispersion around the average $\mu(z,t)$ over the transverse directions shows that the fluid has reached approximately a local (metastable) equilibrium in the plane $x-y$.}
\label{Fig8}
\end{center}
\end{figure}

It is a priori unclear whether this quantity has a true physical meaning but it turns out that  the $\mu_i$'s  at a given $z$ fluctuate very little from site to site, as illustrated in Fig. \ref{Fig8} by a series of histograms collected at different times. One can see in Fig. \ref{Fig7}  that $\mu(z,t)$ ahead of the imbibition front is initially much lower than $\mu_{sat}$ due to the attraction exerted by the solid (this corresponds to the initial redistribution of the vapor inside the aerogel noticed above). The effective chemical potential then grows with time as the density of the wetting film and the gas density in the largest cavities both increase.

\begin{figure}[hbt]
\begin{center}
\includegraphics[width=9cm]{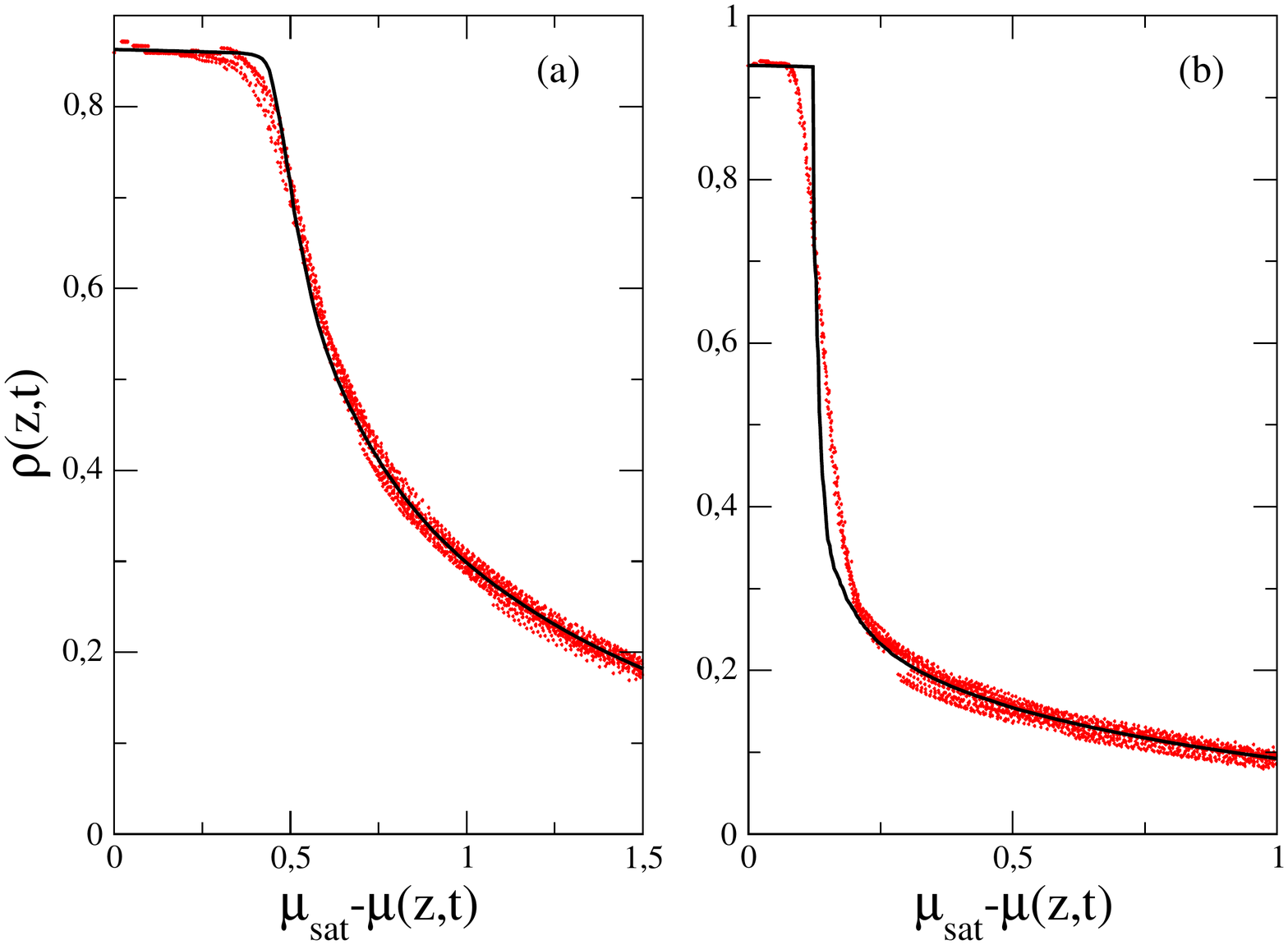}
 \caption{(Color on line) Dynamic density profiles $\rho(z,t)$ in the $87$\% (a) and $95$\% (b) aerogels at different times as a function of the chemical potentials $\mu(z,t)$ (red points): comparison with the corresponding gas adsorption isotherms $\rho(\mu)$ (black solid curves). Note that the fluid density is here plotted as a function of  $\mu_{sat} -\mu(z,t)$ (resp. $\mu_{sat} -\mu$) instead of $\mu(z,t)$ (resp. $\mu$). }
\label{Fig9}
\end{center}
\end{figure}

Now, plotting  $\rho(z,t)$ as a function of $\mu_{sat}-\mu(z,t)$, as done in Fig. \ref{Fig9}, reveals that the density profiles at different imbibition times collapse on a {\it single} master curve which is very close (though not fully identical) to the corresponding adsorption isotherm. This surprising finding strongly suggests that the same fluid configurations (or at least very similar configurations) are encountered during the imbibition and adsorption processes. This can be checked by comparing  the configurations obtained dynamically at time $t$ to those obtained adiabatically   {\it in the same plane} $z$ via a grand-canonical adsorption protocol with $\mu\equiv \mu(z,t)$. This comparison (not shown here) proves unambiguously that the same states are indeed visited in the $87\%$ aerogel whereas some differences occur in the $95\%$ one\cite{note4a} [whence the small but noteworthy differences between the two curves in the vertical part of the isotherm plotted in Fig. \ref{Fig9} (b)].

\begin{figure}[hbt]
\begin{center}
\includegraphics[width=8cm]{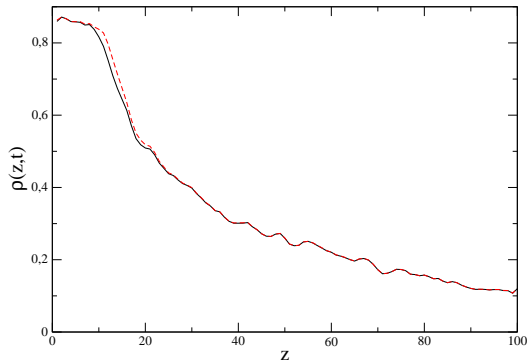}
 \caption{(Color on line) Comparison between the imbibition profile in a $87\%$ porosity aerogel at $t=60$ (solid black line) and the density profile obtained  from Eqs. (\ref{Eqmin}) using the fictitious external potential $V_{ext}(t,z)=\mu_{sat}-\mu(t,z)$ and initial conditions $\rho_i=0$ (red dashed line).}
\label{Fig10}
\end{center}
\end{figure}

How can one rationalize this remarkable result ? Firstly we note that the quasi-uniformity of the $\mu_i$'s in the direction perpendicular to the flow\cite{note4} implies that the flux $J_{ij}(t)$ from a site $i$ to a nearest-neighbor site $j$ in the same plane is approximately zero (indeed, inserting the definition of $\mu_i$ in  Eqs. (4) and (5) allows one to rewrite Eq. (\ref{Eqflux})  as $J_{ij}(t)=w_{ij}\rho_i(1-\rho_j)[1-\exp(-\beta(\mu_i-\mu_j))]$\cite{M2008}). In other words, the fluid ``equilibrates" much more rapidly in $x-y$ plane than in the direction of the flow (here, ``equilibrates" means that some local minimum in the energy landscape has been reached). This becomes more and more accurate as the imbibition front  advances and slows down, as illustrated by the decrease of the variance in the histograms of Fig. \ref{Fig8}. Secondly, as can be seen in Fig. \ref{Fig7}, $\mu(z)$ varies very slowly with $z$ and remains essentially constant over all characteristic lengths in the system (in particular, the aerogel correlation length $\xi_G$; see the discussion below).  Therefore, at time $t$, the fluid structure at the distance $z$ (or, say, in a slice ``around" $z$) is the one observed in the same slice of a {\it three-dimensional} metastable state, {\it i.e.} a local minimum of the grand potential $\Omega$ for $\mu=\mu(z,t)$. There are actually many metastable states for a given value of $\mu$ (visited when performing incomplete adsorption/desorption cycles\cite{KMRST2001,DKRT2003}), but since $\mu(z,t)$ inside the aerogel is very negative at the beginning of the imbibition process, it is not surprising that the states visited successively during the process are (to a good approximation) the same extremal states as in an adsorption experiment. As a further evidence of this assertion, we compare in Fig. \ref{Fig10} the imbibition profile at time $t=60$ with the (static)  density profile obtained by solving Eqs. (\ref{Eqmin}) with the fictitious nonuniform external potential $V_{ext}(t,z)=\mu_{sat}-\mu(t,z)$, starting from the initial conditions $\rho_i=0$, $\forall i$. Apart from small deviations in the region of the imbibition front, one observes a remarkable agreement between the two curves. 

The above reasoning, however, only applies when the adsorption isotherm is smooth in the thermodynamic limit.  This is always the case in the $87\%$ porosity aerogel because the available empty space between the silica particles is small so that the elementary condensation events (the so-called ``avalanches") remain localized\cite{DKRT2003,DKRT2004,DKRT2005}. On the other hand, in very light aerogels and at low temperature ({\it e.g.}, $\phi=95\%$ and $k_BT/w_{ff}=0.9$), a genuine discontinuity is predicted to occur in the isotherm in the thermodynamic limit at a certain value of $\mu$\cite{note6}. This jump is due to the condensation of the gas inside a large void space spanning the whole porous solid (in other words, the occurrence of a {\it macroscopic} avalanche). The nature and properties of this nonequilibrium,  disorder-induced phase transition  are discussed in detail in Ref. \onlinecite{DKRT2005}. The presence of a discontinuity in the isotherm also signals that there are no available metastable states  (within a certain range of density) in which the system can locally relax. Therefore, within this range of density, the fluid configurations visited during the imbibition process  cannot be genuine metastable states.  This explains the deviations observed in Fig. \ref{Fig9} (b) in the vicinity of the jump in the isotherm, which also corresponds to the range of densities associated to the imbibition front. On the other hand, there is good agreement between the two curves for $\rho \lesssim 0.3$.

 In the preceding discussion, for instance in the parametric plot $\rho(z,t)$ {\it vs.} $\mu(z,t)$ shown in Fig. \ref{Fig9}, $t$ dropped out of the description. We now consider the time evolution of the imbibition profiles. Since imbibition proceeds in several stages that can be more or less distinguished depending on the gel porosity (advance of the precursor film, swelling of the film and filling of the voids between the silica strands, advance of the main front), it is instructive to study these stages separately although they take place simultaneously in the different regions of the solid. 
\begin{figure}[hbt]
\begin{center}
\includegraphics[width=9cm]{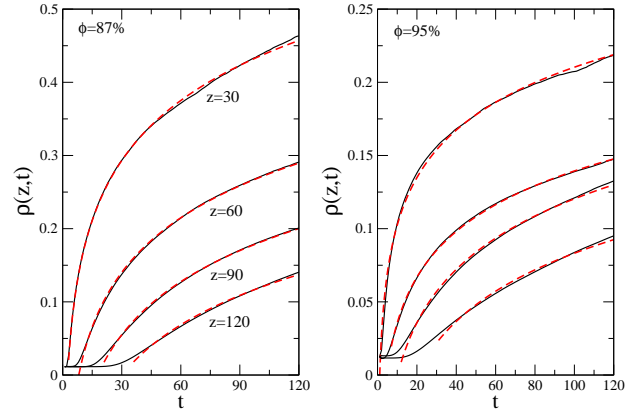}
 \caption{ (Color on line) Time evolution of the average fluid density $\rho(z,t)$ at different heights $z$ in aerogels of  $87$\% (left) and $95$\% (right) porosity (the results have been averaged respectively over three and five gel realizations). The dashed lines correspond to the logarithmic fit $\rho(z,t)=a+b \ln(z/t^{1/2})$ with $a\approx 0.669$ (resp. $0.334$) and $b\approx -0.220$ (resp. $-0.10$) for the $87$\% (resp. $95$\%) aerogel.}
\label{Fig11}
\end{center}
\end{figure}

In Fig. \ref{Fig11} we show the time evolution of the average density $\rho(z,t)$ at different heights above the reservoir {\it before} the arrival of the main front. We first estimate the time $t_{pre}(z)$ that it takes for $\rho(z,t)$  to increase by $10$\% from its initial gas-like value, which we (rather arbitrarily) choose as the signature of the arrival of the precursor film. In both aerogels, this time varies like $z^2$ (for instance $t_{pre}(z) \approx z^2/1000$ in the $95$\% aerogel) indicating that the rise of the wetting film is purely diffusive. This behavior was already observed in the slit pore\cite{KLRT2010} in agreement with molecular and lattice-Boltzmann simulations\cite{DMB2007,C2008}, and is also found in several experiments\cite{BQ2003,DC2010,AAEFD2008}. The density $\rho(z,t)$ then slowly increases with time as the wetting film along the silica strands thickens, the gas density within the largest voids increases and the liquid invades the small crevices between the strands\cite{note5}. As shown in Fig. \ref{Fig11}, the numerical data in this intermediate time regime are well fitted by a logarithmic growth law (although a power-law dependence with a very small exponent cannot be discarded). We have no definite explanation for this behavior but it may simply be due to the use of a nearest-neighbor solid-fluid interaction. It is indeed reminiscent of the logarithmic growth of wetting layers that is found in the case of short-range ({\it i.e.} exponentially decreasing) interface potential\cite{L1985,C1991}. More important in the present context is the observation that $\rho(z,t)\approx a+b\ln(z/t^{1/2})$ with $a$ and $b$ only depending on the porosity. This shows that the growth of the invading liquid domain follows a purely diffusive scaling law in this time regime, despite the complexity of the gel microstructure.

Eventually, the main imbibition front arrives at the height $z$ and the average density  increases up to $\rho_l (1-\phi)$, either smoothly for $\phi=87$\% or more abruptly for $\phi=95$\%. However, even in the lightest aerogel, the front is not sharp.  Therefore, to analyze its time evolution, we extract from the density profiles in Fig. 3 a series of rise level curves corresponding to the different heights $z_{\rho}(t)$ at which the density $\rho$ is reached for the first time (with $0.5\le \rho \le 0.8$ for $\phi=87\%$ and $0.4\le \rho \le 0.8$ for $\phi=95\%$). As shown in Fig. 12,  these curves are almost linear when plotted against $t^{1/2}$, although the data are somewhat noisy due to the small number of gel realizations.  Accordingly, a reasonable collapse of the imbibition fronts at different times is obtained in Fig. \ref{Fig13} by using the scaling variable $z/t^{1/2}$.
\begin{figure}[hbt]
\begin{center}
\includegraphics[width=8cm]{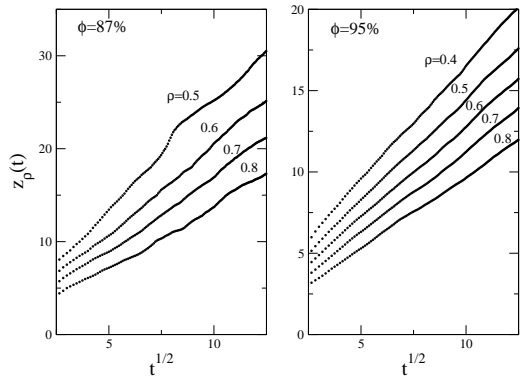}
 \caption{Rise level curves in the $87\%$ and $95\%$ aerogels as a function of $t^{1/2}$. These curves represent the time evolution of the position $z_{\rho}(t)$ of the imbibition front defined by $\rho(z_{\rho}(t),t)=\rho$.}
\label{Fig12}
\end{center}
\end{figure}
\begin{figure}[hbt]
\begin{center}
\includegraphics[width=8cm]{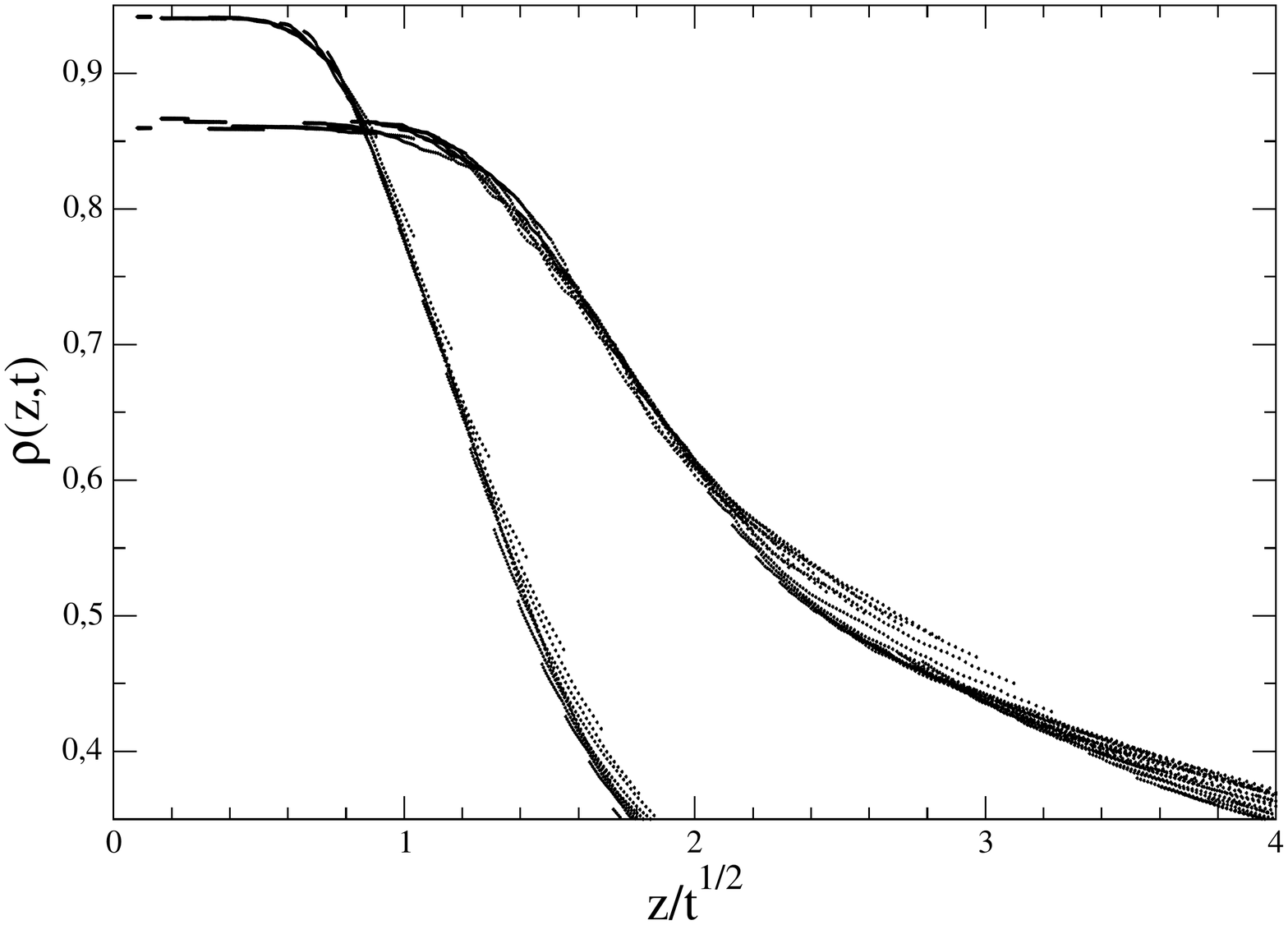}
 \caption{ Average density profiles $\rho(z,t)$ in the $87$\% and $95$\% aerogels plotted as a function of the scaling variable $z/t^{1/2}$. The figure focuses on the imbibition front in the time interval $60\le t\le 150$.}
\label{Fig13}
\end{center}
\end{figure}

\begin{figure}[hbt]
\begin{center}
\includegraphics[width=8cm]{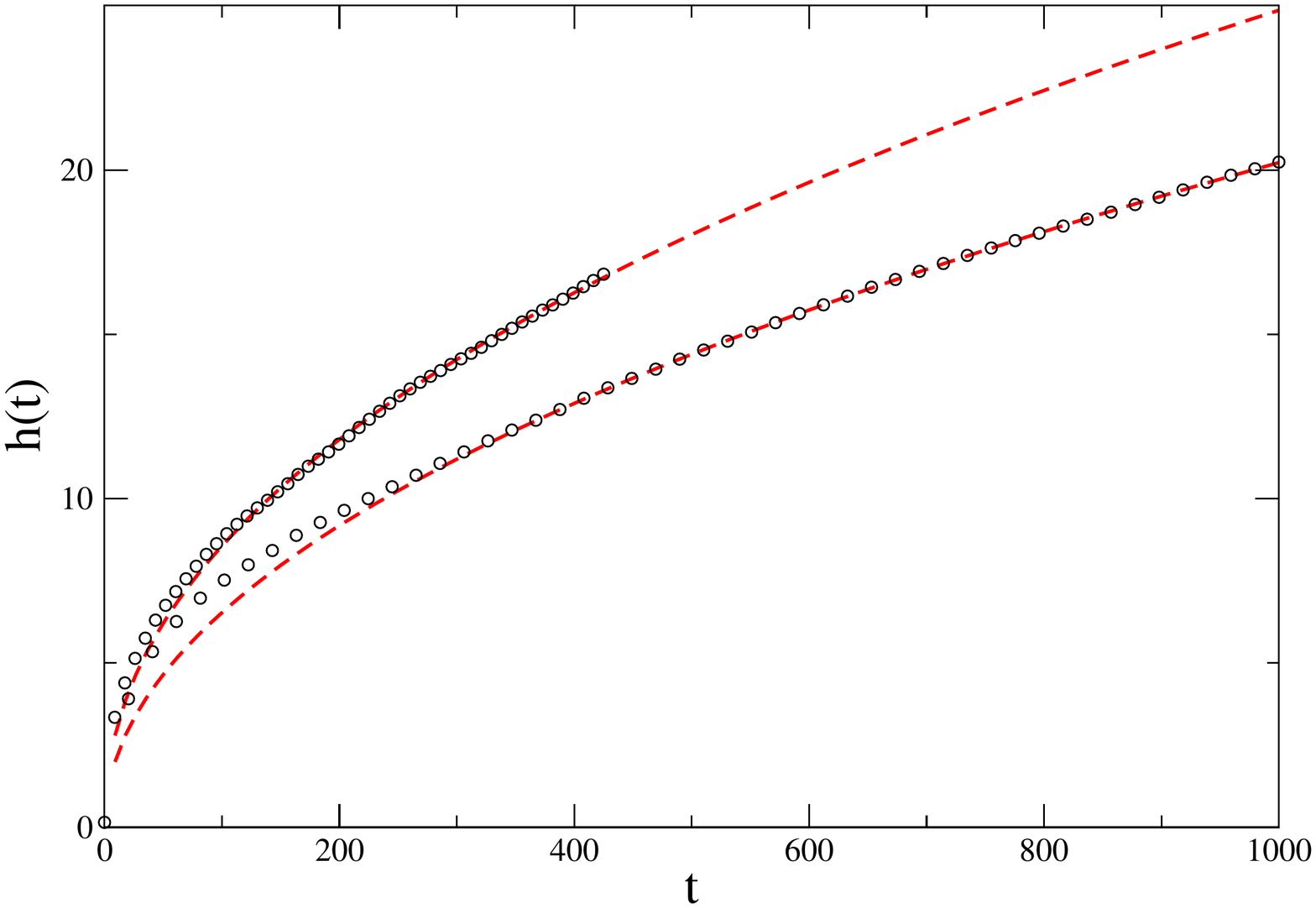}
\end{center}
\caption{(Color on line) Time evolution of the position $h(t)$ of the imbibition front defined by Eq. (\ref{EqH1}) with $\rho_{min}=0.5$ and $\rho_{max}=0.8$. The red dashed lines represent the best fit to the power law $h(t)\sim t^{\alpha}$ with $\alpha=0.49$ for $\phi=87\%$ (bottom curve) and $\alpha=0.46$ for $\phi=95\%$ (top curve). The imbibition times are renormalized here by the porosity-dependent factor $\kappa(\phi)$ given in Table 1. } 
\label{Fig14} 
\end{figure}

From this analysis, it is tempting to conclude that the advance of the imbibition front obeys the Lucas-Washburn $\sqrt{t}$ law for all porosities.  However, when looking more closely, it seems that  the advance of the front is a bit slower in the $95\%$ porosity aerogel. This is illustrated in Fig. \ref{Fig14}  that shows the time evolution of the average position of the front which we define as
\begin{equation}
\label{EqH1}
h(t)=\frac{1}{\rho_{max}-\rho_{min}}\int_{\rho_{min}}^{\rho_{max}} z_{\rho}(t) d\rho
\end{equation}
with $\rho_{min}=0.5$ and $\rho_{max}=0.8$ (these values are not critical as long as $\rho_{min}$ is large enough to leave aside the contribution of the precursor films).  It appears that a better fit of the data for $\phi=95\%$ is obtained with $h(t)\sim t^{0.46}$ (and indeed a slightly better collapse is also obtained in Fig. 13 by using the scaling variable $z/t^{0.46}$). Note that the imbibition times in Fig. 14 are renormalized by the factor $\kappa(\phi)$ to describe the variation of the permeability with porosity.

Considering the limited amount of data at our disposal (due to the very slow progression of the imbibition front in the numerical calculations), it is unclear whether this sub-diffusive behavior actually persists at longer times. However, this may be a genuine signature of the fractality of the aerogel. Although $95$\% aerogels displays fractal character on a rather limited length scale (and moreover the void space filled by the liquid is not fractal), it is plausible that the progression of the imbibition front, which corresponds to the filling of the largest cavities in the gel, is actually sensible to this characteristic feature of the microstructure. However, to settle this delicate issue, it would be necessary to study larger samples and even lighter aerogels, which is computationally too demanding at the moment. 
As shown in Fig 15, the total fluid uptake $\Gamma(t)=\int [\rho(z,t)-\rho(z,0)] dz$ shows a similar behavior although the exponent $\alpha$ is a bit larger as the contribution of the precursor films (which behave purely diffusely in all aerogels) is included.
\begin{figure}[hbt]
\begin{center}
\includegraphics[width=8cm]{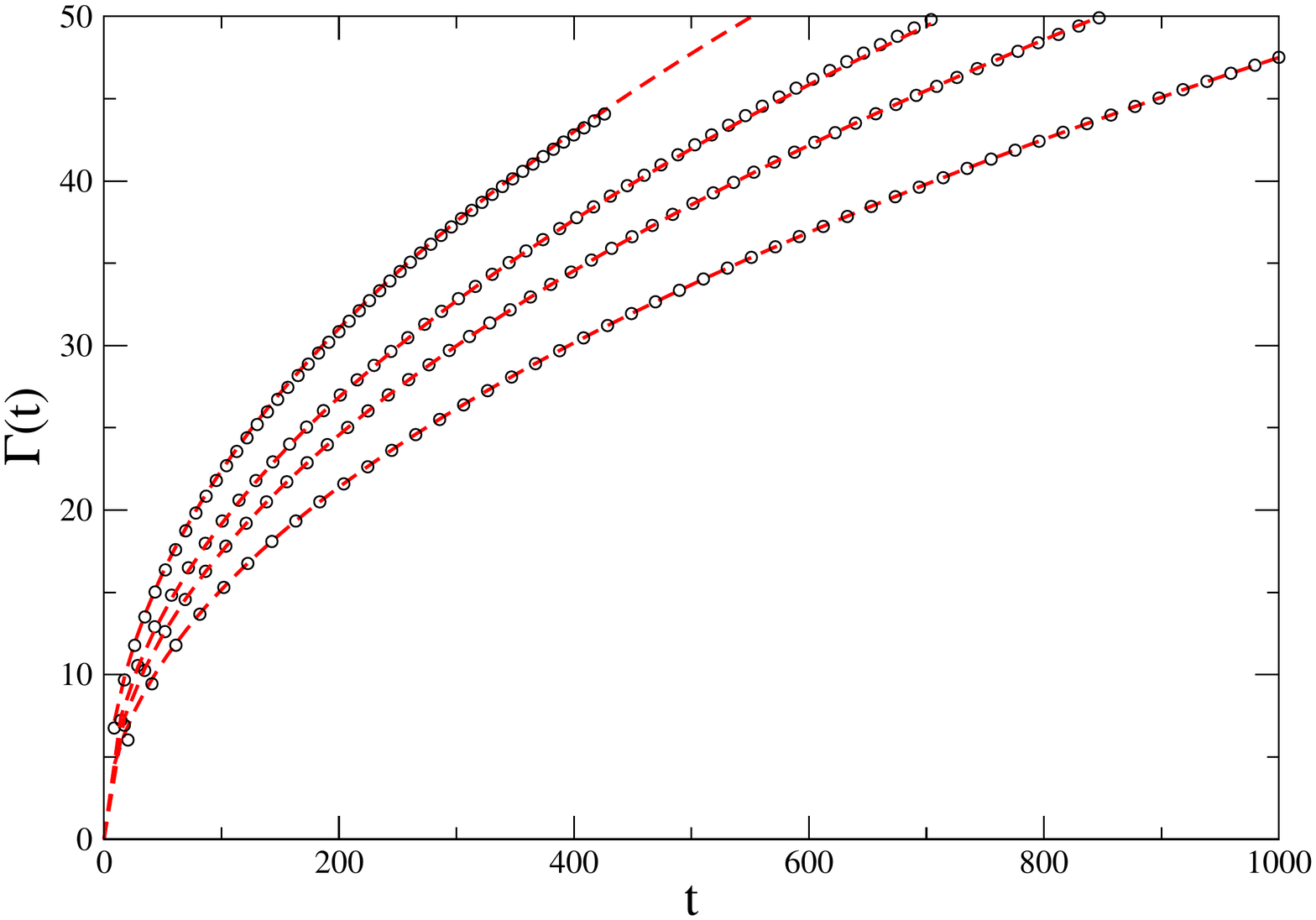}
\end{center}
\caption{(Color on line) Time evolution of the total fluid uptake $\Gamma(t)$. From bottom to top: $\phi=87\%$, $90\%$, $92\%$, and $95\%$. The exponents of the best power-law fit (red dashed lines) are $0.50$, $0.49$, $0.49$ and $0.47$, respectively. The imbibition times are renormalized by the porosity-dependent factor $\kappa(\phi)$ given in Table 1. } 
\label{Fig15} 
\end{figure}

The results displayed in Figs. 12 and 13 also imply that the effective width $W(t)$ of the imbibition front increases as a power of time with an exponent $\beta$ close to $1/2$ (and no smaller than $0.45$ for $\phi=95\%$). Care must be exerted  in drawing definite conclusions from the present calculations\cite{note6} but they are clearly in disagreement with the behavior predicted by phase-field calculations\cite{DREAMA2000} which suggest $\beta\approx 0.3$. In this respect, it must be emphasized that the disordered matrix in phase-field models is described  by quenched, spatially uncorrelated random fields. At this level of coarse-graining, there is no length scale associated to the solid microstructure (for instance, an average pore size $R_p$) and the growth of interface fluctuations is only limited by $L$, the lateral size of the system. These fluctuations are then correlated up to a distance $\xi$ which turns out to be directly related to the progression of the interface so that it grows with time like $t^{1/4}$. Since the interface is found to be super-rough with a global roughness exponent $\chi\approx 1.25$,  this eventually leads to $W(t)\sim t^{0.3}$. It is not clear, however, whether this line of arguments remains valid when the disorder is strong and even the largest pores in the solid are small (in a $87\%$ DLCA aerogel for instance, there are essentially no cavities with a size larger than $5a$, where $a$ is the lattice unit; the gel correlation length $\xi_G$, which is closely related to the average size of the cavities, is of the same order\cite{DKRT2003}).  In such a situation, the actual correlation length in the confined liquid saturates at a value corresponding to the size of the cavities\cite{note7}  and one rather expect the ``trivial'' value $\beta=1/2$, as if the solid were just a bundle of independent pores. In fact, this is the value found in recent experiments performed with water and hydrocarbons in porous Vycor glass\cite{G2010}, a result that was interpreted as signaling the absence of network effects during the imbibition process. Confinement effects are indeed strong in Vycor which has a low porosity ($\phi\approx 30\%$) and a small average pore radius ($R_p\approx 3-5$ nm).  It is possible, but remains at the speculative level, that there is a critical disorder (in the present case, a critical value of the porosity) separating a ``strong-disorder" regime dominated by confinement, with trivial $\beta=1/2$ and a ``weak-disorder" regime where the asymptotic $\beta$ is close to 0.3. 

Finally, it remains to compare the time evolution of the imbibition process in the different aerogels. This requires to renormalize the time for taking into account the influence of the porosity on the permeability $\kappa(\phi)$, as discussed in section 2. The results displayed in Fig. 14  show that the imbibition front rises faster in the lightest aerogel, which is indeed what is observed in experiments\cite{N2006}. The same behavior is found for the total fluid uptake $\Gamma(t)$ displayed in Fig. 15. As discussed in the Appendix, these results are essentially unchanged when taking into account local variations of the permeability.

\section{Conclusion}

In this paper, we have presented a theoretical study of spontaneous imbibition of helium in silica aerogels using a coarse-grained lattice model of the gel microstructure and a mean-field treatment of the fluid dynamics. Our main approximation, which should be valid in the case of a slowly advancing front, is that capillary disorder predominates over permeability disorder, as also assumed in phase-field treatments of spontaneous imbibition.  To our knowledge, this work is the first three-dimensional study of imbibition in a disordered porous solid.  The main findings can be summarized as follows:

1) Irrespective of porosity, the first stage of imbibition corresponds to the advance of a liquid film along the silica strands and in the small crevices of the microstructure. The gas density then increases in the larger voids between the strands while the precursor film thickens. The largest cavities in the aerogel eventually fill with liquid, which may be associated to the passage of the imbibition front. It is however difficult to clearly distinguish the front from the precursor film except in the case of the lightest gels ({\it e.g.}, a $95\%$ porosity aerogel).

 2) The time evolution of the whole process is in general purely diffusive but there are some indications that the advance of the front could be slower than $\sqrt{t}$ in very light aerogels, which may be related to the fractal character of the solid.
 When taking into account the dependence of the `effective' permeability on the porosity, one finds that the front advances faster in light aerogels, as observed experimentally.
 
3) The effective width of the interface also varies approximately like $\sqrt{t}$, which is not the behavior predicted by phase-field models with uncorrelated random fields.  Although our results are too limited to draw definite conclusions, this indicates that wetting and confinement modify the interface roughening.

4) Due to the quasi-static behavior of the imbibition process, the fluid configurations observed during imbibition and gas adsorption bear a strong resemblance. As a consequence,  when parametrically plotting the average fluid density against the average chemical potential at a fixed  height and at various imbibition times, one obtains a master curve that almost identifies with an adsorption isotherm computed in the grand canonical ensemble. This correspondance breaks down in high-porosity aerogels at low temperature when a jump discontinuity is present in the adsorption isotherm, signaling a first-order, nonequilibrium, disorder-induced phase transition.

These findings are in partial agreement with the experimental  observations of Ref. \onlinecite{N2006} with aerogels of  porosities $90.4, 95.8$ and $99.5\%$. There is a clear change in the fluid behavior as the porosity increases, which is induced by the presence of very large cavities. This can be put in parallel with the change of behavior observed in adsorption isotherms, as correctly suggested by the authors of Ref. \onlinecite{N2006}. Moreover, it is plausible that the distinction between the precursor film and the main imbibition front is enhanced visually by the non-linearity of the eye vision, and can be interpreted as the existence of two distinct advancing interfaces. However, even in the case of the lightest aerogel, we find that the first interface is always advancing faster than the main front, contrary to what has been observed experimentally.  We thus rather believe that the two advancing interfaces seen in the experiments correspond to either a unique interface viewed from different directions or to two distinct imbibition fronts generated by macroscopic inhomogeneities in the aerogel.

\acknowledgments
This work was supported by ANR-06-BLAN-0098.

\appendix

\section{Using a local permeability}

In this Appendix we propose and test a modification of the theoretical treatment that takes into account the local environment of the fluid particles inside the gel in order to introduce some disorder in the  permeability $\kappa$. The idea is to define a local porosity at each site by taking into account the effect of the environment within a given distance. More specifically, we introduced a coarse-grained porosity field $\phi_i(r_0)$ which is obtained from the mean gel density in a sphere of radius $r_0$ around site $i$. The local permeability at site $i$ is then obtained from Brinkman's expression,
\begin{align}
\label{A1}
\kappa_i(r_0)&=\frac{2R^2}{9(1-\phi_i(r_0))}\nonumber\\
& \times
\left[1+\frac{3}{4}(1-\phi_i(r_0))\left(1-\sqrt{\frac{8}{1-\phi_i(r_0)}-3}\right) \right] .
\end{align}

\begin{figure}[hbt]
\begin{center}
\includegraphics[clip=true,width=8cm]{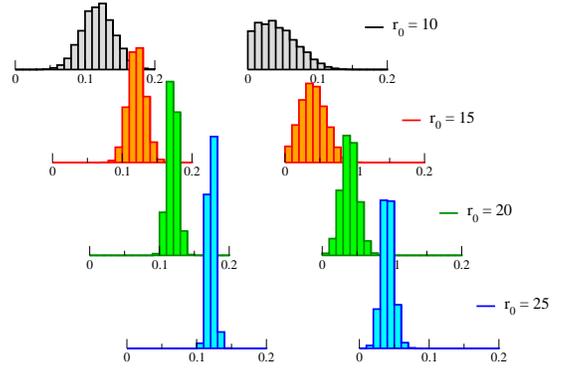}
\end{center}
\caption{Histograms of the averaged solid density $1-\phi_i(r_0)$ for several values of $r_0$ in the $87\%$ (left) and $95\%$ (right) aerogels.} 
\label{Fig16} 
\end{figure}

A natural choice for $r_0$  is the aerogel correlation length $\xi_G$ (or the average cluster size estimated in Ref. \onlinecite{DKRT2003} from the location of the first minimum of the two-point correlation function) but, to illustrate the effect of $r_0$, we report  in Fig. \ref{Fig16} the histograms of the averaged solid densities  $1-\phi_i(r_0)$ for several values of $r_0$ : for each site $i$, we compute $\phi_i(r_0)=\sum_{j}\eta_j/\sum_{j}1$ where the sum is performed over sites $j$ whose distance from site $i$ is less than or equal to $r_0$;
then,  one builds histograms from all the empty sites.  For aerogels of porosity 
$\phi=87\%$ and $\phi=95\%$, one has $\xi_G\simeq 4a$ and $\xi_G\simeq 10a$, respectively \cite{DKRT2003}. As expected, when $r_0$ is too small compared to $\xi_G$,
 there is a non-zero probability that the coarse-grained density is zero,  which is not acceptable since the corresponding Brinkman's permeability then diverges (see Eq. \ref{A1}). On the other hand, when $r_0$ becomes larger, the distribution is more peaked around the mean permeability of the material. Taking $r_0$ equal to $\xi_G$ avoids the spurious effect of a local permeability that diverges and captures reasonably well the local morphology ({\it i.e.}, the correlations) of the medium.  
 
Finally, we have to introduce the local permeability into the evolution equation. This can be done in different ways provided that the conservation of mass is still verified and the equilibrium density distribution does not change.  These conditions are satisfied with the following functional form of the evolution equation :
\begin{equation}
\label{evolB}
\frac{\partial\rho_i}{\partial t}=-\sum_{j/i}\kappa_{ij}J_{ij}(t)=-\sum_{j/i}J'_{ij}(t).
\end{equation}
where $J_{ij}$ is still defined by Eq. \ref{Eqflux} and the $\kappa_{ij}$'s  are coefficients of a permeability matrix which has to be explicited. 
The condition on mass conservation can be written as $J'_{ij}(t)=-J'_{ji}(t)$, which implies $\kappa_{ij}=\kappa_{ji}$: therefore the permeability matrix must be symmetric. For simplicity, we have chosen $\kappa_{ij}=[\kappa_i(r_0=\xi_G)+\kappa_j(r_0=\xi_G)]/2$ where $\kappa_i$ is defined in Eq. (\ref{A1}).

It turns out that the fluid density profiles so obtained differ very little from those computed with $\phi_i=\phi$. The only deviations occur in the density range corresponding to the main imbibition front and associated to the passage of the liquid in the largest voids of the aerogel. These voids fill up more rapidly because the local porosity in the middle of the cavities reaches high values. As a result, the front advances a little bit faster. Overall, the differences are however negligible. One cannot exclude that they would increase as time increases, but the procedure using local permeabilities is very time-consuming and we have not been able to simulate longer times.


\begin{thebibliography}{10}
\bibitem{note1} The aerogels are initially generated in a cubic box of lateral size $L$ with periodic boundary conditions in all directions; in order to generate a gel of height $L_z=2L$, we simply duplicate the original system using a mirror symmetry at $z=L$. This procedure ensures that the resulting gel is connected and is harmless for the imbibition process because fluid configurations at $z=L$ are in any case quite different from those at $z=0$. The same procedure is repeated to generate a gel of height $L_z=4L$.
\bibitem{note1a} In principle, saddle points are also solutions, but being unstable they are not accessible in standard numerical resolutions.
\bibitem{note2} Note that the bulk lattice gas has a symmetric coexistence curve $\rho_l(T) = 1 - \rho_g(T)$ with chemical potential $\mu_{sat}=-4w_{ff}$ in the mean-field approximation. The fact that $\mu_{sat}$ is independent of T is due to the artificial hole-particle symmetry of the bulk lattice gas.
\bibitem{note3} The lateral size $L$ is also important since it eventually controls the fluctuations of the interface width\cite{ADR2004}, but we do not consider this issue in the present work.
\bibitem{note4a} In the $95\%$ aerogel at the distance $z=50$, the liquid configurations for $t<100$ are also identical during the two processes because the imbibition front has not yet arrived as can be seen in Fig. \ref{Fig3}.
\bibitem{note5} Note that because of the disordered nature of the solid, we do not observe in aerogels successive layering transitions as was the case in a single slit pore\cite{KLRT2010}.
\end{thebibliography}
\end{document}